# A compact single-shot soft X-ray photon spectrometer for free-electron laser diagnostics

Kirk A. Larsen,[1,2] Kurtis Borne,[4] Razib Obaid,[1] Andrei Kamalov,[1] Yusong Liu,[1] Xinxin Cheng,[1] Justin James,[1] Taran Driver,[1] Kenan Li,[1] Yanwei Liu,[1] Anne Sakdinawat,[1] Christian David,[3] Thomas J. A. Wolf,[1,2] James Cryan,[1,2] Peter Walter,*[1,5] and Ming-Fu Lin*[1]

**Affiliations**:
1. SLAC National Accelerator Laboratory, Menlo Park, CA, USA
2. Stanford PULSE Institute, SLAC National Accelerator Laboratory, Menlo Park, CA, USA
3. Paul Scherrer Institute, CH-5232 Villigen PSI, Switzerland
4. Department of Physics, Kansa State University, Manhattan, KS, USA
5. TAU Systems, Austin TX, USA

**Abstract**:
The photon spectrum from free-electron laser (FEL) light sources offers valuable information in time-resolved experiments and machine optimization in the spectral and temporal domains. We have developed a compact single-shot photon spectrometer to diagnose soft X-ray spectra. The spectrometer consists of an array of off-axis Fresnel zone plates (FZP) that act as transmission-imaging gratings, a Ce:YAG scintillator, and a microscope objective to image the scintillation target onto a two-dimensional imaging detector. This spectrometer operates in an energy range which covers absorption edges associated with several atomic constituents: carbon, nitrogen, oxygen, and neon. The spectrometer's performance is demonstrated at a repetition rate of 120 Hz, but our detection scheme can be easily extended to 200 kHz spectral collection by employing a fast complementary metal oxide semiconductor (CMOS) line-scan camera to detect the light from the scintillator. This compact photon spectrometer provides an opportunity for monitoring the spectrum downstream of an endstation in a limited space environment with sub-electronvolt energy resolution.

## 1. Introduction

A photon spectrometer is a crucial asset to any X-ray beamline. Monitoring the broadband photon spectrum from stochastic free-electron laser (FEL) light sources provides essential feedback to optimize the performance of the FEL, such as for sub-femtosecond pulse duration for studying attosecond electron dynamics and nonlinear optics [1–3]. Collecting the spectra downstream of the interaction point (IP) provides additional information in time-resolved experiments making use of the recently developed spectral-correlation spectroscopy [4,5] and optical density (OD) measurements. Our photon spectrometer is particularly convenient for transient absorption spectroscopy that compares the transmitted FEL spectra with and without optical excitations (i.e., ΔOD measurement), circumventing time-consuming fine scans in photon energy. This is similar to the transient absorption measurement using tabletop high-harmonic generated (HHG) X-ray light sources. In contrast to HHG sources, FEL light sources have much higher pulse energies that allow single-shot spectrum collection, which is essential for the correlation measurements with electrons and ions on a shot-by-shot basis. Accurate measurements of the X-ray photon spectrum is needed to help with selective and unambiguous excitation of the resonant features in the target or to provide reference features to assign to the time-dependent signals in the pump-probe experiments. However, the photon energy calculated from machine parameters (e.g., the electron beam kinetic energy and the undulator gaps of FEL) frequently deviates from the true value by a few electronvolts (eV) to a few tens of eV, so a fast and reliable energy calibration of the soft X-ray spectrum and its central photon energy are basic experimental requirements.

Here, we show that by using an off-axis Fresnel zone plate (FZP), a Ce:YAG scintillator (Princeton Scientific Corp.) and a microscope (Navitar Inc.), we are able to measure single-shot photon spectra at X-ray energies in the neighborhood of the carbon, nitrogen, oxygen, and neon K-edge absorption features at sub-eV energy resolution. A similar approach has been demonstrated at the SwissFEL facility to measure the spectrum at hard X-ray photon energies [6]. The alignment of the X-ray laser through the off-axis FZP is simple [7], and the diffracted spectrum along the optical axis is measured at the maximum rate of our current available camera. This was sufficient for a single-shot characterization at the 120 Hz repetition rate at which we performed our measurements. The characteristic line-shaped spectrum with a width of a few pixels perpendicular to the spectral axis suggests that a fast line-scan complementary metal oxide semiconductor camera (CMOS) can be used for the future ~100s kHz repetition rate operation of LCLS-II. The compact size of this spectrometer can be integrated into a limited space in existing beamlines where photon spectrum measurements are required.

## 2. Design of the soft X-ray photon spectrometer

The off-axis FZP gratings were made of gold-patterned thin film on a 100 nm thick silicon nitride membrane ($Si_3N_4$) for the carbon, oxygen, and neon absorption K-edges. For the nitrogen edge, we replaced the $Si_3N_4$ membrane with a 100 nm aluminum membrane because of the strong absorption of $Si_3N_4$ at this photon energy. As a plating base, we evaporated 5 nm Cr and 10 nm Au on membranes. The FZP gratings were then patterned using e-beam lithography on PMMA resist, followed by gold electroplating to ~100 nm. Table 1 shows the design parameters for the FZP gold patterning on top of the substrates. Here, the pitch distance represents the spacing of the outermost zone. The smaller pitch for neon is used to increase the dispersion, however, for the carbon absorption edge a larger pitch is used to reduce the dispersion so that in all cases, the Ce:YAG detector plane is kept at a similar distance away from the X-ray main axis (12 to 17 mm from the zero-order main beam). Each FZP grating has a certain bandwidth that operates in a reasonable energy resolution. The energy resolutions are estimated and measured at several X-ray energy ranges, and the results are shown in section 3.

**Table 1:** Pitch and radius for the FZP at the outermost edge with the designed focal distance of 441 mm. All the grating architecture on top of the substrate are made of gold material, while the substrates vary with the absorption edges. The pixel-to-eV factor is used to convert CCD pixel number to electronvolt (e.g., 69 pixels per eV near 290 eV). The full energy range is calculated using the full 1024 pixels of the OPAL-CCD sensor. The resolution limit of FZP is calculated from the limited size of the FZP aperture of 0.5x0.5 $mm^2$ [7].

| Elements | Energy (eV) | Pitch (nm) | Radius (mm) | substrate | Pixel-to-eV factor | Full Energy (eV) | Resolution limit $E_{fzp}$ (eV) |
|---|---|---|---|---|---|---|---|
| **carbon** | 290 | 150 | 12.574 | $Si_3N_4$ | 69 | 14.8 | 0.087 |
| **nitrogen** | 410 | 100 | 13.341 | Aluminum | 49 | 20.9 | 0.082 |
| **oxygen** | 530 | 100 | 10.319 | $Si_3N_4$ | 29 | 35.3 | 0.106 |
| **neon** | 870 | 70 | 8.928 | $Si_3N_4$ | 17 | 60.2 | 0.122 |

Figure 1 shows the schematic of the photon spectrometer design as implemented in the TMO hutch at the LCLS [8,9]. The schematic includes the upstream existing Kirkpatrick–Baez (KB) mirrors. The nominal focal length of the horizontal (vertical)-focusing KB mirror is 2.4 m (1.6 m). The source point of the FZP is located at the IP, where a gas needle effusive source, a pulsed molecular beam, a gas flow cell, or a solid-state thin film can be used as sample targets. The focal length of the KB mirrors can be adjusted within a range of ±100 mm from the nominal focus. The minimum focal spot size at the IP is ~1.3x1.1 µm$^2$ measured by our wavefront sensor targets using the fractional Talbot effect from a grating [9,10]. As the comparable size of the horizontal and vertical foci implies, we can use the width along the non-dispersive direction to calculate the resolution along the spectral axis. The designed distance from the IP to the FZP is approximately 1.801 m ($g_0$). This distance is greater than the designed distance between the FZP and the Ce:YAG detector (~0.584 m, $b_0$). This provides a magnification (M) of the focal spot at the IP, down by a factor of ~0.32 at the Ce:YAG detector plane (i.e., focal spot of ~0.42x0.35 µm$^2$). The main advantage of this demagnification is to reduce the stretching length of spectra on the detector due to the chromatic aberration of the FZP while scanning a large energy window, for instance, ~50 eV at 870 eV photon energy (see Equation 1 and the blue line on the detector in Figure 1). This is particularly convenient for a large energy range scan without moving the Ce:YAG detector inside the vacuum chamber. An energy calibration curve (i.e., the pixel-to-energy mapping and absolute energy offset) can be obtained readily from scanning the FEL photon energy over a broad range while the detector position and the associated microscope imaging system are fixed.

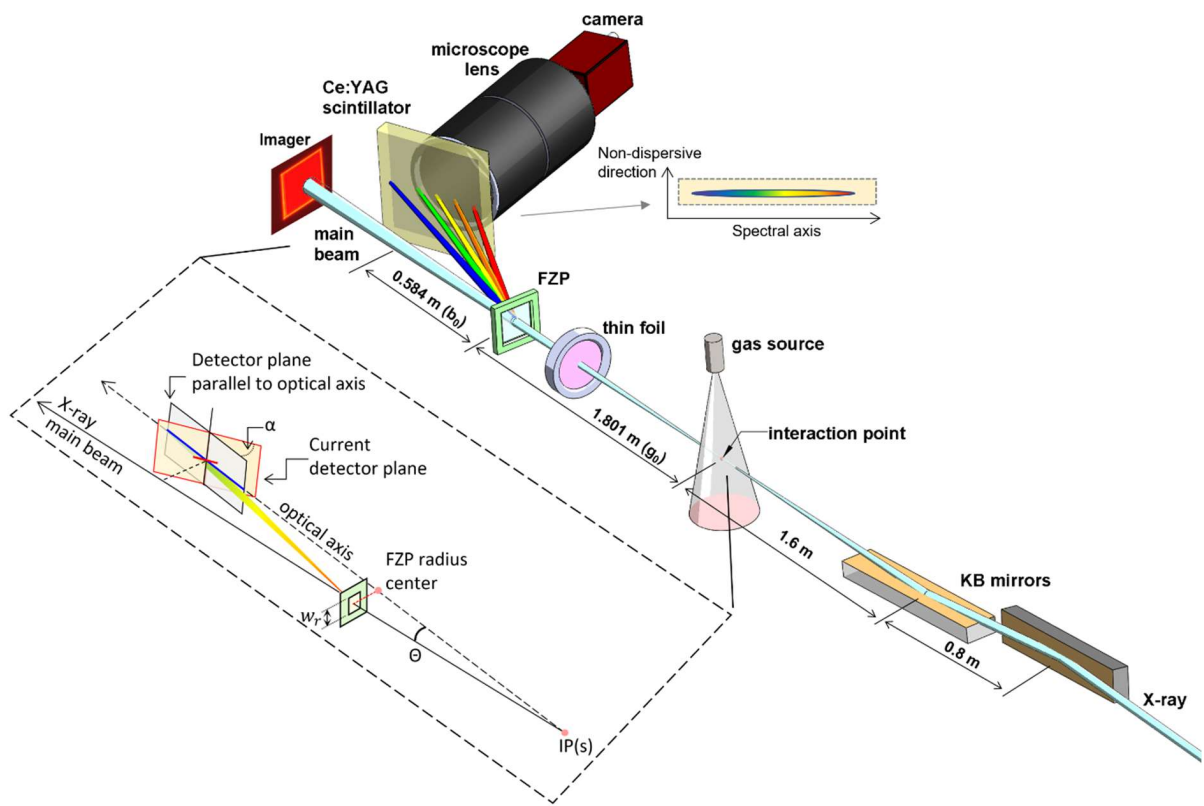

**Figure 1**. Schematic of the photon spectrometer and upstream Kirkpatrick–Baez (KB) mirrors. The X-ray laser beam propagates toward the left. The interaction point (IP) is the position of the nominal focal point of the KB mirrors and this point is adjustable with a range of ± 100 mm. The designed distance from the IP to the FZP is 1.801 m (i.e., $g_0$) and the distance from FZP to the center of the Ce:YAG is approximately 0.584 m (i.e., $b_0$). A thin foil is used for the X-ray energy

calibration. A gas source highlights the IP location, which is the source point of the spectrometer. The microscope contains a high zoom lens and a 2D CCD camera that runs at the full rate of LCLS. The distance between the front of the microscope lens to the front surface of Ce:YAG is ~32 mm. Note that the microscope lens and camera are located outside the chamber, and the rest of components are in vacuum. The transmitted zero-order main beam through the FZP is projected onto the downstream 2D detector (imager) for alignment purposes. A spectral region of interest is also shown to highlight the spectral axis and non-dispersive direction of a single-shot spectrum. The dimension of the spectrum along the non-dispersive direction can be used to estimate the energy resolution. A layout that describes the X-ray main beam, optical axis and stretching length of spectrum at two different detector angles are shown on the bottom left.

The relation of chromatic aberration, $\Delta b$, of the FZP with the X-ray energy E and bandwidth $\Delta E$ is shown in Equation 1 [7].

$$\Delta b = f(1 + M)^2 \frac{\Delta E}{2E} \quad \text{(Eq. 1)}$$

$$\frac{1}{f} = \frac{1}{b_0} + \frac{1}{g_0} \quad \text{(Eq. 2)}$$

, where f is the focal length of the FZP, and M is the magnification of the spectrometer. The magnification is defined as $b_0/g_0$, corresponding to 0.32 in the current design. Using Equation 2, this gives a focal length of 441 mm for an FZP at the designed imaging distances and photon energies as shown in Table 1. The demagnification (M<1) of the focal spot implies the detector system should be able to resolve structure at a few-micron scale to enable the highest energy resolution on the detector surface. The Ce:YAG scintillators have been demonstrated to be high quality X-ray detectors, showing < 1 µm parallax and fluorescence spreading [11]. Although the quantum efficiency is not high for soft X-ray measurements using Ce:YAG, the spatial resolution is the most important consideration when designing the energy resolution of such a device. From our estimations, considering the transmission efficiency of the FZP, a few µJ of pulse energy can provide a sufficient number of photons on the scintillator for a single-shot spectral measurement. In our studies here, we have demonstrated that ~5 uJ is sufficient to provide a single-shot spectrum at 530 eV, however, several of our experiments using the FZP spectrometer in parallel with a photoelectron/ion measurements have been recording up to ~200 µJ without any sign of degradation in resolution over time or membrane rupture due to thermal heating (i.e., average power of ~0.02 W).

The overall magnification of the microscope lens is set to 1.64x in the current experiments. The pixel size of the CCD is 5.5 µm, therefore each pixel on the CCD sensor represent 3.3 µm on the Ce:YAG (i.e., 5.5/1.64 = 3.3 µm), and the field-of-view (FOV) on the Ce:YAG is ~3.4 mm (with a 5.6 mm CCD sensor size, 5.6/1.64 = 3.4 mm). The low magnification of the microscope lens is to maintain the large FOV so that a wide X-ray energy range can be recorded without moving the Ce:YAG detector plane. At such low magnification, the microscope spatial resolution is limited to ~4.9 µm which corresponds to ~1.5 pixels on the CCD sensor. Note that at the full magnification of the camera lens (i.e., full zoom 9x) a ~1.8 um spatial resolution can be achieved, which will be critical for the future upgrade of the photon spectrometer. In this condition, a larger camera sensor with a spectral axis near ~20 mm is necessary to maintain the FOV for a wide X-ray energy range without moving the detector plane.

We can estimate the signal level of the X-ray photons on our spectrometer system. Since the thickness of the Ce:YAG scintillator (~ 0.5 mm) is much larger than the attenuation length (hundreds of nanometers) of the soft X-rays from 280 eV to 2 keV, all the X-rays impinging on the scintillator are absorbed [12]. The fluorescence spectrum of the scintillator from the X-ray absorption is centered at ~550 nm with a bandwidth of ~100 nm [13]. Each absorbed X-ray photon emits 8 to 16 ~550 nm photons from the Ce:YAG scintillator per keV of X-ray energy [12]. To estimate a lower bound for the signal level, we assume four fluorescence photons per 530 eV photon absorbed. The X-ray beam is approximately 3 mm (FWHM), which is substantially larger than the open aperture of the FZP window (~0.5x0.5 mm$^2$). Cropping of the beam by the FZP reduces the effective number of photons to ~2.4 % of the incident beam, assuming the window is centered at a gaussian profile of X-ray. In addition, the thickness of the $Si_3N_4$ substrate attenuates the X-rays down to 66% due to absorption. The grating efficiency is assumed to be ~10% [7,14]. Therefore, the overall fluorescence yield of the diffracted X-rays illuminated on the scintillator is ~6.3E-3 at 530 eV. The collection efficiency of our microscope is 0.1% using a numerical aperture of 0.066, and our current camera sensor detection efficiency at 550 nm is ~ 40%. Thus, the overall detection efficiency per incoming X-ray photon through the FZP is ~2.5E-6. Assuming the entire dispersive spectrum is spread to 300 pixels, the signal level of each pixel is ~8.5E-9 (i.e., spectral axis of 150 pixels by non-dispersive direction of 2 pixels, see Figure 1). For a pulse energy of ~1 µJ at the photon energy of 530 eV on the FZP grating, this corresponds to 1.2E10 photons per pulse. This suggests that on average we have ~100 electrons on each camera pixel (assuming 1 electron per each 550 nm photon detected). This is a factor of ~8 over the readout noise of our current camera. Thus, at least a few µJ of X-ray pulse energy is sufficient for single-shot spectral measurements using this combination of Ce:YAG and microscope CCD camera. From this estimation, a camera with low readout noise is an important parameter to increase the spectrometer detection sensitivity. Our current OPAL camera (Adimec OPAL-1000) has a noise level of 14 electrons per pixel per readout at the full repetition rate of 120 Hz (i.e., current LCLS-I full rate).

## 3. Single-shot spectra, energy resolution and discussion

We used our spectrometer to measure the single-shot spectrum near the carbon, nitrogen, oxygen, and neon K-edges, as illustrated in Figure 2. Averages of the single-shot measurements are shown above the single-shot data. The estimated X-ray beam size which illuminates on the FZP array is photon energy-dependent, and ranges from ~1 to 5 mm (from 860 eV toward 280 eV). Therefore, the total pulse energy varies from ~20 µJ to ~100 µJ, in order to keep the same fluence (i.e., a few mJ/cm$^2$) on the FZP grating to avoid damages and maintain similar signal levels. These pulse energies provide sufficient signal-to-noise levels to record single-shot spectra at 120 Hz. For each FZP, we can resolve the SASE sub-structure of the individual X-ray pulses, in addition to the shot-to-shot jitter in the X-ray central photon energy ($\hbar\omega_c$). The average spectrum suggests a ~1 % FWHM bandwidth from the FEL (i.e., $0.01\hbar\omega_c$). Note that the 2D spectra (a1, b1, c1 and d1) are not plotted in a 1-to-1 aspect ratio. The width along the vertical axis (non-dispersive direction) is an image of the source point (i.e., the IP) and this value can be used to estimate the energy resolution for each FZP.

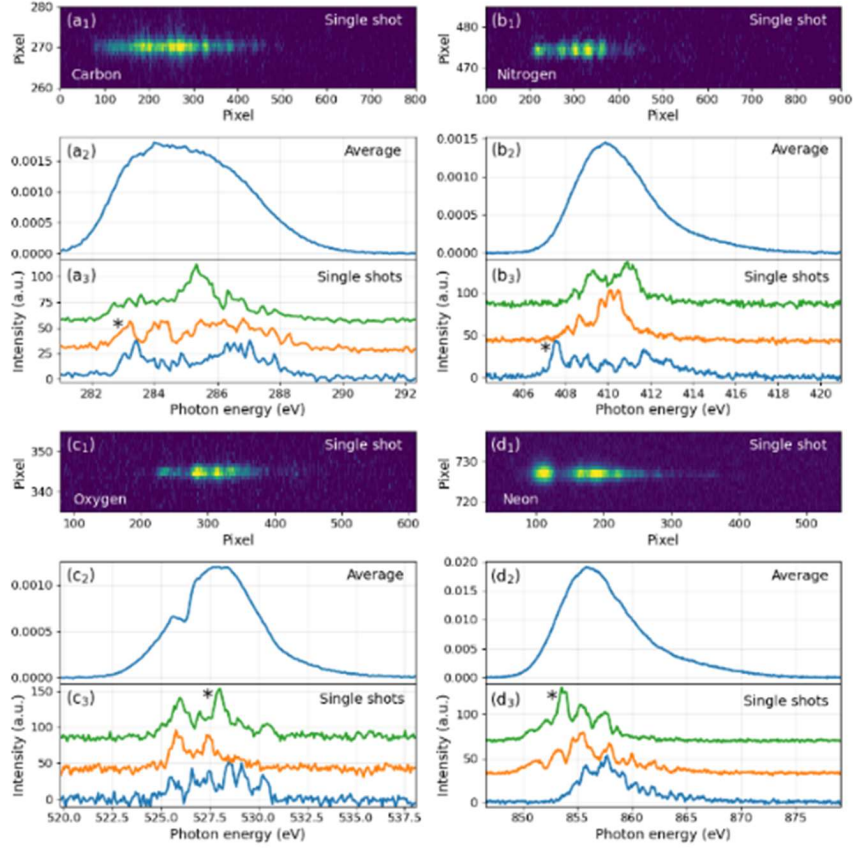

**Figure 2.** Single-shot and average spectra of the FZPs at designed photon energies near the carbon, nitrogen, oxygen, and neon absorption K-edges. The pulse energy on the FZP varies from ~20 uJ to ~100 uJ. The higher X-ray photon energy requires lower pulse energy because of a smaller spatial spot on the FZP grating. (a) Carbon K-edge spectra (b) Nitrogen K-edge spectra (c) Oxygen K-edge spectra and (d) Neon K-edge spectra. The small dip on the averaged spectrum in (C) can be from the oxidized layer of mirror coating along the beamline or the absorption of the Ce:YAG detector that contains oxygen element. Note that the single-shot image shown in (a1, b1, c1 and d1) are not plotted in a 1-to-1 aspect ratio. The widths of the spectra in FWHM along the non-dispersive (vertical) direction are only a few pixels (~2 to 5 pixels). The asterisks exemplify SASE sub-structures in the single-shot spectra in a3, b3, c3 and d3.

The spectrometer is calibrated with thin metal foils upstream of the FZP grating (between the IP and the FZP as displayed in Figure 1). The absorption spectra of these target foils are obtained by comparing the averaged transmission spectra with and without the samples in the beam path. The absorbance (i.e., OD) of the sample is calculated using

$$OD = \log_{10}\left(\frac{I_0}{I}\right) \quad \text{(Eq. 3)}$$

, where I and $I_0$ are the transmission spectra with and without the sample, respectively. The absorption was measured from four different targets: a 0.5 μm carbon film, a 0.9 μm mylar film, a thin CuO film and a 0.15 μm nickel film. The carbon, mylar and nickel thin films are all purchased from Lebow Company. The CuO film is made by oxidation of a ~70 nm Cu film (deposited on 100 nm $Si_3N_4$) in an air furnace at ~600 °C for 2 hours. We also measured the absorption through a nitrogen gas column (~0.07 torr in a ~14 meters long gas attenuator) which is upstream of the TMO endstation. The measured optical density using the FZP (red) is plotted in comparison with measurements from literature (blue) in Figure 3(b)-3(f). The transmission spectra with (green) and without (grey) carbon film are shown in Figure 3(a) to demonstrate the OD calculation. The ~1% bandwidth of the FEL average spectrum can cover the entire near-edge resonance features of gaseous and solid systems, which is an advantage for energy calibration during the experiments. Note that the resonant absorption features may not be located at the position of the best resolution in the designed FZP due to the tilted angle of the detector plane relative to the optical axis. However, the resolved SASE sub-structure features suggest that sub-eV energy resolution is achieved across these broad photon energy ranges. In addition, the absorption widths in semiconductors or metals do not represent the actual resolution since the width of the absorption peaks is a convolution between the core-hole lifetime and the broad transitions from the core-level electron to the conduction bands. The sharp peaks seen in the nitrogen gas, however, provide the direct resolution measurement of ~0.21 eV.

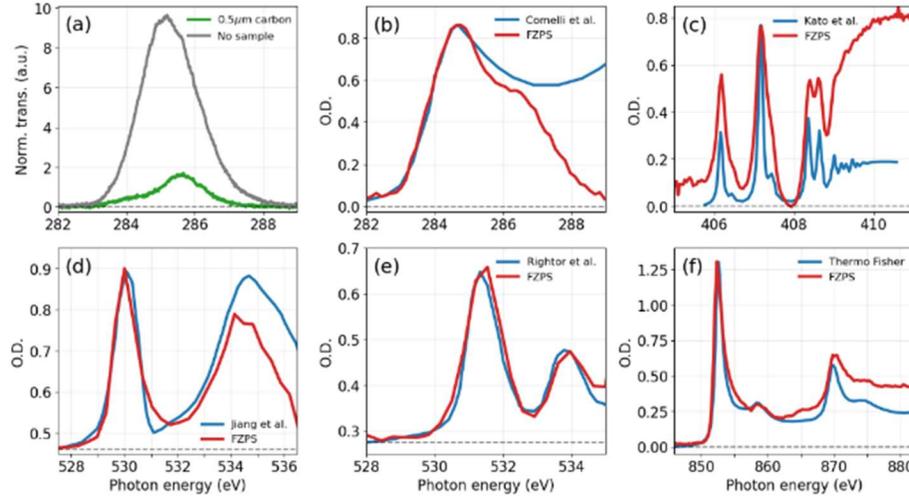

**Figure 3.** (a-b) Photoabsorption spectrum measurement at the carbon K-edge using a carbon film (0.5 μm), (c) nitrogen K-edge using nitrogen gas, (d-e) oxygen K edge using CuO and mylar (0.9 μm) thin films and (f) absorption region near neon K-edges using a nickel thin film (0.15 μm) calibration, respectively. The transmission spectra with and without samples are shown in (a) for the carbon K-edge absorption as an example of the optical density calculation using Eq. 3. The absorption spectra shown in (b-f) are measured by the FZP, and are displayed in red, while reference spectra are shown in blue [15–18]. The peaks observed in the $N_2$ absorption spectrum originate from N 1s → Rydberg state excitations [15].

Two measurements are performed to quantify the energy resolution of the spectrometer. The first measurement was scanning the source point of the X-ray and monitoring the vertical width change of the single-shot spectrum. The second measurement was a scan of photon energy and was performed to find the optimal energy resolution at the current tilted detector plane (~9 degree relative to the optical axis). Both measurements used the widths along the non-dispersive direction of the spectrum. This width is an image of the source point, s, from the KB foci at the IP (see Figure 1), and this axis provides information on the maximum resolution achievable with a FZP based spectrometer. Equations 4-10 below are used to calculate the final energy resolution of a FZP at the current geometry of the photon spectrometer design. In Equation 4, the $\Delta x_0$ represents the nominal X-ray focal size on the detector with the source ($g_0$) and image ($b_0$) distances at the designed values.

$$\Delta x_0 = s\left(\frac{b_0}{g_0}\right) \qquad \text{(Eq. 4)}$$

The image size on the Ce:YAG detector, $\Delta x_1$, is found by combining the nominal image size with two broadening contributions as shown in Equation 5. The first part is the size of the nominal image size due to demagnification of the source point. The second part is the width when the KB focus is at the distance upstream or downstream of the IP ($g \neq g_0$ and $b \neq b_0$). The $\Delta \bar{b}$ is defined as ($b-b_0$), which describes the focal distance offset of the FZP due to the change in the KB foci away from the designed value $g_0$. The third part is the spot size broadening due to the chromatic aberration [7]. This provides the magnitude of the spot broadening due to the energy offset between the photon energy and the designed X-ray energy of a specific FZP (i.e., $\Delta b \neq 0$ when $\Delta E \neq 0$). $\Delta b$ is the chromatic aberration described in Equation 1. The $w_r$ is the open aperture of the FZP (see Figure 1).

$$\Delta x_1 = \sqrt{(\Delta x_0)^2 + (\Delta \bar{b} \frac{w_r}{b})^2 + (\Delta b \frac{w_r}{b})^2} \qquad \text{(Eq. 5)}$$

The Δx in Equation 6 represents the final focal-spot width on the camera sensor with the consideration of the current microscope resolution, $\Delta x_2$ (~4.9 µm), and the width change, $\Delta x_1$, as described in Equation 5.

$$\Delta x = \sqrt{\Delta x_1^2 + \Delta x_2^2} \ ; \Delta x_2 = 4.9 \ \mu m \qquad \text{(Eq. 6)}$$

The spatial broadening described up until in Equation 6 is complete for the estimation of the expected width along the non-dispersive direction for comparison to data.

The spatial widths along the non-dispersive direction from Equation 6 and from experimental measurements are both converted to units of energy by considering two factors: (1) the spreading of the diffracted light along the detector plane due to the detector's tilt and (2) the pixel-to-eV factor. The stretch width (i.e., footprint) on the tilted detector is calculated using Equation 7,

$$\Delta x' \approx \Delta x / \sin(\theta + \alpha) \qquad \text{(Eq. 7)}$$

where θ is the diffraction angle between the zero-order X-ray and the optical axis, and $\alpha$ is the tilted angle of the detector relative to the optical axis (see Figure 1). The angle θ and α are approximately ~0.33 and 9 degrees in the current design for 530 eV.

The pixel-to-eV factor (pte), shown in Equation 8, can be obtained from the energy scan while recording the central pixel of the average spectrum (see Table 1).

$$\Delta E' = \Delta x' / pte \qquad \text{(Eq. 8)}$$

Due to the limited size of the FZP aperture, the energy resolution is restricted to ~0.1 eV (i.e., $E_{fzp}$) as calculated from Equation 9 [7]. This value is further converted to the final resolution using the geometric mean shown in Equation 10. The calculated energy resolution (as a function of KB focal distances, photon energies and FZP choices) is compared to experiments in Figures 4 and 5.

$$E_{fzp} \approx \frac{pitch}{w_r} E \qquad \text{(Eq. 9)}$$

$$\Delta E = \sqrt{\Delta E'^2 + E_{fzp}^2} \qquad \text{(Eq. 10)}$$

In the first measurements, we adjust the KB focal length near the designed interaction/source point to locate the minimum width in the non-dispersive direction so that the spectral resolution can be optimized for each designed FZP. The widths at several locations away from the designed focal point are recorded. Figure 4 displays the predicted results as dot-dashed and solid curves, respectively, for Δb=0 and Δb≠0 (i.e., ΔE=0 and ΔE≠0). The measured values are shown as circles with error bars for three different photon energies. The measured minima in Figure 4 are higher than the dot-dashed predicted values by ~20% to 50 %. This suggests that the X-ray energies used in the measurement are a few eV different from the X-ray energies for which the FZPs were designed. This explanation is supported by using sharp resonant absorption features of nitrogen gas, where we measured a resolution of ~0.21 eV at 406 eV, shown as the red square in Figure 4(a). The difference between the two measurements using nitrogen absorption and the width of the non-dispersive direction indeed originates from the strong variation in the focal width as a function of X-ray energy at the tilted detector (i.e., 409 eV for the foci scan and 406 eV for the absorption measurement), which is supported by the estimation using ~409 eV as shown as red solid line. This also indicates that an energy difference of ~1 % at the nitrogen K-edge has an impact on the energy resolution. Similarly, the discrepancy between measurements and dot-dashed prediction in Figures 4(b-c) can also be explained by difference of energies. A small energy offset in the estimation, shown in solid lines, can describe experimental results better (i.e., 3 eV and 2 eV for oxygen and neon cases). The effect of the tilted detector that leads to resolution loss is shown in Figure 5.

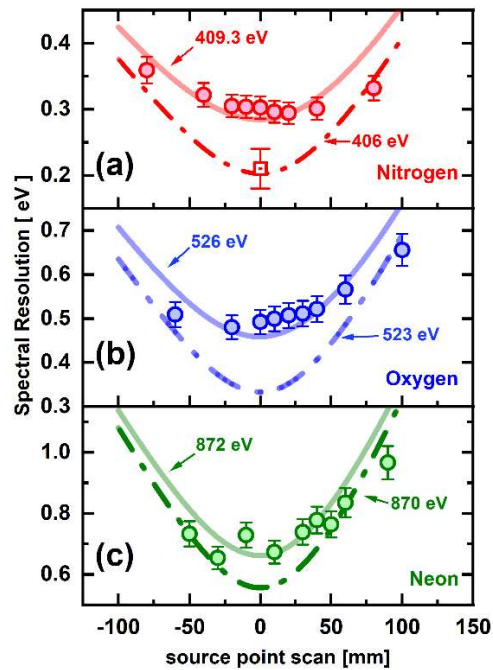

**Figure 4.** The spectral resolution measurements as a function of KB focal distance for three photon energies at the (a) nitrogen (409 eV), (b) oxygen (526 eV) and (c) neon (872 eV) K-edges (red, blue, and green circles, respectively). The resolution is calculated using the FWHM width along the non-dispersive direction (the y-axis of the camera images shown in Figure 2), and those widths are converted to eV. The dot-dashed (solid) lines are the estimated resolution dependence using Equations 4-10 with Δb=0 (Δb≠0). The resolution difference between dot-dashed and solid lines originates from the energy difference ($\Delta b$ and $\Delta E \neq 0$). The error bars represent 0.5 degree uncertainty in the angle of the Ce:YAG detector. The red square in (a) represents the obtained resolution from the nitrogen gas absorption spectrum at 406 eV and it agrees with the prediction shown in the dashed curve.

For the second experiment, we scanned the photon energy to test the resolution. Although tilting the detector encompasses a larger X-ray energy range within the CCD sensor, the energy resolution degrades when the photon energy deviates from the designed X-ray photon energy (see Table 1). In principle, this can be mitigated by reducing the tilt angle between the detector plane and the optical axis in order to simultaneously achieve a good focal condition for a broad X-ray energy range. However, this requires the whole detector plane to lie nearly parallel to the optical axis, and it inevitably leads to a long image length in the spectral axis and covers a very small energy range by the limited CCD sensor size. In the current spectrometer design, the tilt angle is approximately 9 degrees, which is a compromise to achieve a ~2% energy range, $0.02\hbar\omega_c$, that gives a resolving power of >1000. Figure 5 shows the resolution variation of the FZPs near the oxygen and neon K-edges at the source distance $g_0$ ($\Delta\bar{b}\sim 0$). Similarly, we again use the width along the non-dispersive direction to estimate the resolution as a function of a broad X-ray energy range. The solid lines are the predicted values using Equations 4-10. The predictions and measurements agree well with each other. The best resolution for the current oxygen FZP is approximately 0.45 eV at 524 eV which is slightly higher than the X-ray photon energy at 526 eV in Figure 4(b). As described above, this explains, in part, the discrepancy between the measured minimum value and prediction in Figure 4(b). This also suggests that a ~1% energy deviation from the designed photon energy leads to a ~10% degradation in energy resolution.

The current limitations on energy resolution are the microscope spatial resolution, tilted angle of the detector and the aperture size of FZP windows. These limitations can be mitigated by incorporating a line-scan camera that is ~20 mm along the spectral axis and ~20 µm along the non-dispersive direction. A longer 1D sensor enables a higher magnification value and smaller tilted angle while maintaining the same FOV, improving the energy resolution of the spectrometer. In addition, increasing the FZP window size from the current 0.5x0.5 mm$^2$ to 1.0x1.0 mm$^2$ will also increase the resolution, while also increasing the risk of film rupture due to mechanical vibration and air flow during the vacuum pump down or venting process.

With a tilted detector in the current FZP spectrometer, the stretching length of the spectrum is significantly reduced (see red and blue lines on the detector in Figure 1) so a larger X-ray energy range can be covered within a given CCD sensor. This is especially convenient for an energy scan with a fixed detector location so the experimental apparatus can remain untouched. However, this inevitably reduces the energy resolution when the incident X-ray energy is set away from the designed photon energy. By reducing the tilt angle from 9 degrees to 4.5 degrees, our FZP can cover ~4% X-ray bandwidth ($0.04\hbar\omega_c$) while still providing a resolving power of better than 1000. The ~4% window should also cover all the important absorption features in static and pump-probe experiments. For instance, the valence hole states for gas or solid systems are generally ~1 to 2 % below the absorption pre-edges [19,20]. These transient features can be measured at high resolution when the right FZP is chosen. For a larger bandwidth scan, the constant optimal resolution can be achieved by using a combination of FZPs with ~4 % energy difference to cover a wider energy range near that absorption edge. This implies that the FZPs can be fabricated to have energies centered at 500 eV, 520 eV and 540 eV for the measurements near the oxygen K-edge and a resolving power of 1000 or higher can be maintained.

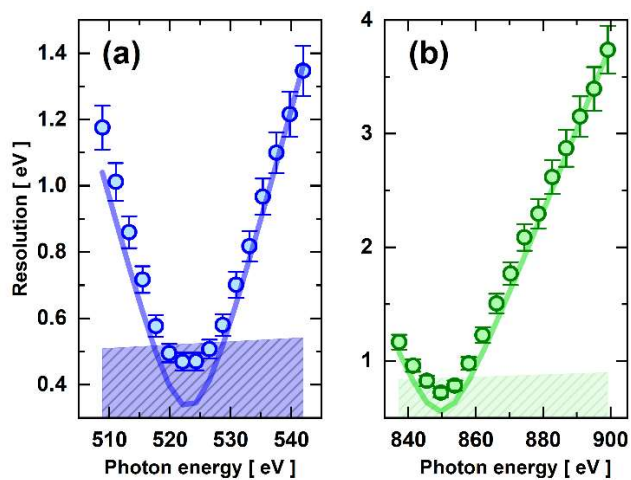

**Figure 5.** Variation of energy resolution for the FZP spectrometer at two different designed photon energies (a) oxygen and (b) Neon K-edges. The energy resolution is calculated from the width of the non-dispersive direction at each photon energy. The predictions shown as solid lines are calculated by considering Equations 4-10. The shaded areas represent a resolving power of 1000 for both the oxygen and neon K edges. The error bars represent 0.5 degree uncertainty in the Ce:YAG tilt angle relative to the optical axis (see Figure 1).

## 4. Conclusion and outlook

We have demonstrated that a FZP can be used to characterize single-shot FEL spectra and demonstrated the energy calibration of the device using multiple sample targets. These measurements suggest sufficient energy resolution to resolve the SASE sub-features of FEL pulses and resonant absorption lines of solid and gas molecular systems, despite the small size of the instrument. This spectrometer has been recently employed by LCLS Users as a spectral diagnostic for FEL tuning and experiments which exploit the correlation between the incident spectrum and measurements made at the IP [21]. We also demonstrated that with the ~2% energy bandwidth of the FEL average spectra (~10 eV bandwidth at 520 eV X-ray), the resolving power can be maintained between 1000 and 1500.

Looking towards the future, the current spectrometer design will require upgrades since the upcoming LCLS-II light source operates at a high repetition rate up to 1 MHz. The average power will be increased by ~3 orders of magnitude to several watts that can easily damage the FZP grating due to the thermal heat deposited on the thin film. In order to mitigate the heat on the FZP, one option is to reduce the pulse energy by 2 to 3 orders of magnitude after the IP and to increase the sensitivity of the detection system. This detection system can be a direct detection such as X-ray detectors or an indirect measurement using the current design of imaging the fluorescence of Ce:YAG by the microscope. However, the direct measurement using a commercial X-ray CCD or MCP has an acquisition rate that does not meet high repetition rate requirements. A custom-made 1-D line camera can have single-photon sensitivity and can operate at high repetition rate. This technical development is currently ongoing at SLAC National Accelerator Laboratory. For indirect detection, by integrating a line-scan camera into the microscope, which has a fast frame rate, and an intensifier, this may provide a good solution for the overall spectrometer operations at the LCLS-II light source.

## Acknowledgement

Use of the Linac Coherent Light Source (LCLS), SLAC National Accelerator Laboratory, is supported by the U.S. Department of Energy, Office of Science, Office of Basic Energy Sciences under Contract No. DE-AC02-76SF00515.